\documentclass[a4paper]{jpconf}
\usepackage{graphicx}

\usepackage{graphicx}
\usepackage{dsfont}     
\usepackage{subeqnarray}

\newcommand{\be}{\begin{equation}}
\newcommand{\ee}{\end{equation}}
\newcommand{\beq}{\begin{eqnarray}}
\newcommand{\eeq}{\end{eqnarray}}

\newcommand{\bsa}{\begin{subeqnarray}}
\newcommand{\esa}{\end{subeqnarray}}

\newcommand{\I}{{_I}}
\newcommand{\II}{{_{I\!I}}}

\def\boldvec#1{\mbox{\boldmath $#1$\unboldmath}}

\def\lsim{\hbox{ \raise.35ex\rlap{$<$}\lower.6ex\hbox{$\sim$}\ }}
\def\gsim{\hbox{ \raise.35ex\rlap{$>$}\lower.6ex\hbox{$\sim$}\ }}

\newcommand{\bea}{\begin{eqnarray}}
\newcommand{\eea}{\end{eqnarray}}
\newcommand{\beaa}{\begin{eqnarray}}
\newcommand{\eeaa}{\end{eqnarray}}
\newcommand{\ba}{\begin{array}}
\newcommand{\ea}{\end{array}}
\newcommand{\bit}{\begin{itemize}}
\newcommand{\eit}{\end{itemize}}
\newcommand{\ben}{\begin{enumerate}}
\newcommand{\een}{\end{enumerate}}

\def\lab{\label}

\def\lf{\left}

\def\pa{\partial}
\def\ran{\rangle}

\def\ri{\right}

\def\al{\alpha}

\def\ga{\gamma}
\def\Ga{\Gamma}

\def\Om{\Omega}
\begin{document}

\noindent KCL-PH-TH/2012-{\bf 6}

\title{Noncommutative spectral geometry, dissipation and the origin of
quantization}

\author{Mairi Sakellariadou} \address{Department of Physics, King's
  College, University of London, Strand WC2R 2LS, London, U.K.}
\ead{mairi.sakellariadou@kcl.ac.uk} \author{Antonio Stabile}
\address{Facolt\`a di Scienze and Gruppo Collegato I.N.F.N.,
  Universit\`a di Salerno, I-84100 Salerno, Italy}
\ead{anstabile@gmail.com} \author{Giuseppe Vitiello}
\address{Facolt\`a di Scienze and Gruppo Collegato I.N.F.N.,
  Universit\`a di Salerno, I-84100 Salerno, Italy}
\ead{vitiello@sa.infn.it}

\begin{abstract}
We present a physical interpretation of the doubling of the algebra,
which is the basic ingredient of the noncommutative spectral geometry,
developed by Connes and collaborators as an approach to unification.
We discuss its connection to dissipation and to the gauge structure of
the theory. We then argue, following 't~Hooft's conjecture, that
noncommutative spectral geometry classical construction carries
implicit in its feature of the doubling of the algebra the seeds of
quantization.
\end{abstract}

\section{Noncommutative Spectral Geometry}
\label{motivation}
Unification of all forces, including gravity, remains one of the open
issues in theoretical physics. As one approaches Planckian energy
scales, the assumption that physics can be described by the sum of the
Einstein-Hilbert and Standard Model (SM) actions breaks down and one
must consider quantum gravity effects. In an attempt to provide a
basis for describing the quantum nature of space-time, which may lead
to the unification of all forces, Connes and collaborators developed
Noncommutative Spectral Geometry (NCSG) which combines notions of
noncommutative geometries~\cite{ncg-book1,ncg-book2} with spectral
triples.  Within NCSG, the SM of electroweak and strong interactions
is seen~\cite{ccm} as a phenomenological model, which specifies the
geometry of space-time so that the Maxwell-Dirac action
functional leads to the SM action.

This unification model lives by construction at high energy scales,
offering an appropriate framework to address early universe
cosmology~\cite{Nelson:2008uy} - \cite{Sakellariadou:2011dk}.
This is however beyond the scope of this presentation. In what follows
we attempt instead, to shed some light on how some criticisms raised
against NCSG approach, and in particular its application in early
universe cosmology, can be answered. More precisely, we will discuss
the physical meaning of the choice of the {\sl almost} commutative
geometry and its relation to quantization~\cite{PRD}. In our discussion we will also consider
the relation of the NCSG formalism with the gauge structure of the theory and with dissipation.

Let us first however review the main elements of NCSG.  It is based on
a two-sheeted space, made from the product of a four-dimensional
smooth compact Riemannian manifold ${\cal M}$ (a continuous geometry
for space-time) with a fixed spin structure, by a discrete
noncommutative space ${\cal F}$ (an internal geometry for the SM)
composed by only two points.  The noncommutative nature of the
discrete space ${\cal F}$ is given by a spectral triple $({\cal A, H,
  D})$, where ${\cal A}$ is an involution of operators on the
finite-dimensional Hilbert space ${\cal H}$ of Euclidean fermions, and
${\cal D}$ is a self-adjoint unbounded operator in ${\cal H}$.  The
space ${\cal H}$ is the Hilbert space $L^2({\cal M},S)$ of square
integrable spinors $S$ on ${\cal M}$ and the algebra ${\cal A}$ is the
algebra ${\cal A}=C^\infty({\cal M})$ of smooth functions on ${\cal
  M}$ and acts in ${\cal H}$ by multiplication operators.  The
operator $D$ is the Dirac operator
${\partial\hspace{-5pt}\slash}_{\cal
  M}=\sqrt{-1}\gamma^\mu\nabla_\mu^s$ on the spin Riemannian manifold
${\cal M}$.

Within NCSG all information about space is encoded in the algebra of
coordinates ${\cal A}$.  Assuming the algebra ${\cal A}$ constructed
in the geometry ${\cal M}\times {\cal F}$ is symplectic-unitary, it
must be of the form~\cite{Chamseddine:2007ia}
\begin{equation}
\mathcal{A}=M_{a}(\mathds{H})\oplus M_{k}(\mathds{C})~,
\end{equation}
with $k=2a$ and $\mathds{H}$ being the algebra of quaternions.  The
field of quaternions $\mathds{H}$ plays an important r\^ole in this
construction and its choice remains to be explained. To obtain the SM
one assumes quaternion linearity.  The first possible value for the
even number $k$ is 2, corresponding to a Hilbert space of four
fermions, but this choice is ruled out from the existence of
quarks. The next possible value is $k=4$ leading to the correct number
of $k^2=16$ fermions in each of the three generations.

The noncommutative spectral geometry model is based upon the spectral
action principle stating that, within the context of a product
noncommutative geometry, the bare bosonic Euclidean action is given by
the trace of the heat kernel associated with the square of the
noncommutative Dirac operator and is of the form
\be {\rm Tr}(f(D/\Lambda))~; \ee
$f$ is a cut-off function, $\Lambda$ fixes the energy scale,
$D$ and $\Lambda$ have physical dimensions of a mass.  This action can
be seen {\sl \`a la} Wilson as the bare action at the mass scale
$\Lambda$.  The fermionic term can be included in the action
functional by adding $(1/2)\langle J\psi,D\psi\rangle$, where $J$ is
the real structure on the spectral triple and $\psi$ is a spinor in
the Hilbert space ${\cal H}$ of the quarks and leptons.

For the four-dimensional Riemannian geometry, the trace ${\rm
  Tr}(f(D/\Lambda))$ is expressed perturbatively in terms of the
geometrical Seeley-deWitt coefficients $a_n$, which are known for any
second order elliptic differential operator,
as~\cite{sdw-coeff,ac1996,ac1997,nonpert}
\beq\label{asymp-exp} {\rm Tr}(f(D/\Lambda))&\sim&
2\Lambda^4f_4a_0+2\Lambda^2f_2a_2+f_0a_4+\cdots\nonumber
\\&& +\Lambda^{-2k}f_{-2k}a_{4+2k}+\cdots~,
\eeq
where  the
smooth even cut-off function $f$, which decays fast at infinity,
appears through its momenta $f_k$ given by:
\beq \nonumber 
f_0 &\equiv& f(0)\\  \nonumber
f_k &\equiv&\int_0^\infty f(u) u^{k-1}{\rm
  d}u\ \ ,\ \ \mbox{for}\ \ k>0 ~,\nonumber\\ \mbox
    f_{-2k}&=&(-1)^k\frac{k!}{(2k)!} f^{(2k)}(0)~.  \nonumber
\eeq
Since its Taylor expansion at zero vanishes, the
asymptotic expansion Eq.~(\ref{asymp-exp}) reduces to
\be {\rm Tr}(f(D/\Lambda))\sim
2\Lambda^4f_4a_0+2\Lambda^2f_2a_2+f_0a_4~.  \ee
The cut-off function $f$ plays a r\^ole only through its three momenta
$f_0, f_2, f_4$, which are three real parameters, related to the
coupling constants at unification, the gravitational constant, and the
cosmological constant, respectively. The term in $\Lambda^4$ gives a
cosmological term, the term in $\Lambda^2$ gives the Einstein-Hilbert
action functional, and the $\Lambda$-independent term yields the
Yang-Mills action for the gauge fields corresponding to the internal
degrees of freedom of the metric.

In this purely geometric approach to the SM, the fermions provide the
Hilbert space of a spectral triple for the algebra, while the bosons
are obtained through inner fluctuations of the Dirac operator of the
product geometry.  The computation of the asymptotic expression for
the spectral action functional results to the full Lagrangian for the
Standard Model minimally coupled to gravity, with neutrino mixing and
Majorana mass terms.

Finally, let us clarify in which sense we
talk of dissipation in what follows. This is
necessary because the Standard Model, as is well known, is 
Quantum Field Theory (QFT) model describing a closed
(nondissipative) system. 
Dissipation enters our discussion of the implications of the algebra
doubling in the specific sense one observes that in 
electrodynamics neither the
energy-momentum tensor of the matter field, nor that of the gauge
field, are conserved.  However, $\partial_{\mu} T^{\mu
  \nu}_{\rm matter} = e F^{\mu \nu} j_{\mu} = - \partial_{\mu} T^{\mu
  \nu}_{\rm gauge\ field}$. Thus, the {\sl total} $T^{\mu \nu}_{\rm
  total} = T^{\mu \nu}_{\rm matter} + T^{\mu \nu}_{\rm gauge
  \ field}$, which is the energy-momentum tensor of the {\it closed}
system \{matter field, electromagnetic field\} is conserved: each element of the
couple is {\sl open} (dissipating) on the other one, although the {\sl
closeness} of the total system is ensured.  In this sense, 
dissipation considerations in our discussion below 
do not spoil the closeness of the SM.

\section{The algebra doubling and the
gauge structure}
\label{classical}

In this Section we study the  relation between the
two-sheeted space in the NCSG construction and the gauge structure of
the theory. 
One central ingredient in NCSG is indeed the ``doubling'' of the
algebra ${\cal A} \, \rightarrow \,{\cal A}_1\otimes {\cal A}_2$
acting on the ``doubled" space ${\cal H}={\cal H}_1\otimes {\cal H}_2$.
Such a doubling  is the formal realization of the NCSG two-sheeted space.
Let us first observe that the doubling of the algebra is also present in the standard Quantum Mechanics (QM) formalism of 
the density matrix and Wigner function. Indeed,  the 
expression of the Wigner function is~\cite{Feynman:1972a}
\bea \lab{W} W(p,x,t) = \frac{1}{2\pi \hbar}\int {\psi^* \left(x -
\frac{1}{2}y,t\right)\psi \left(x + \frac{1}{2}y,t\right)
e^{-i\frac{py}{\hbar}}dy} ~. \nonumber \eea
By putting $x_{\pm}=x\pm \frac{1}{2}y$, 
the associated density matrix is
\be\lab{AA8} W(x_{+},x_{-},t) \equiv \langle x_{+}|\rho (t)|x_{-}\rangle =
\psi^* (x_{-},t)\psi (x_{+},t)~, \ee
The
coordinate $x(t)$ of a quantum particle is thus split into two coordinates
$x_+(t)$ (going forward in time) and $x_-(t)$ (going backward in
time). The forward  and the backward in time evolution of the
density matrix $W(x_{+},x_{-},t)$ is then described by ``two copies"
of the Schr\"odinger equation:
\be i\hbar {\partial \langle x_+|\rho (t)|x_-\rangle \over \partial t}=
H \langle x_+|\rho (t)|x_- \rangle~, \lab{(5a)} \ee
where $H$ is given in terms of the two Hamiltonian operators $H_{\pm}$ as $H \,=\, H_+ -H_-$.
The connection with Alain Connes' discussion of
spectroscopic experiments and the algebra doubling is thus evident: the density
matrix and the Wigner function {\it require} the introduction of a
``doubled" set of coordinates $(x_{\pm}, p_{\pm})$ 
and of their respective algebras. 
Use of
Eq.~(\ref{(5a)}) shows immediately that the eigenvalues
of $H$ are directly the Bohr transition frequencies $h
\nu_{nm}=E_n-E_m$, which was the first  hint towards an explanation
of spectroscopic structure.

The need to double the degrees of freedom is implicit
even in the classical theory when considering the Brownian motion and it is related to dissipation. In the classical Brownian theory one has the equation 
of motion~\cite{Blasone:1998xt}
\be
m\ddot{x}(t)+\ga \dot{x}(t)=f(t)~, \lab{(22)}
\ee
where $f(t)$ is a random (Gaussian distributed) force: $~<f(t)f(t^\prime )>_{\rm noise}=2\,\ga \,k_BT\; \delta (t-t^\prime)$.
By averaging over the
fluctuating force $f$, one obtains~\cite{Blasone:1998xt}
\bea \lab{(25)}
 <\delta[m\ddot{x}+\ga \dot{x}-f]>_{\rm noise}=
\int {\cal D}y <\exp[{i\over \hbar}
\int dt \;L_f(\dot{x},\dot{y},x,y)]>_{\rm noise}~,
\eea
where
\be L_f(\dot{x},\dot{y},x,y)= m\dot{x}\dot{y}+ {\ga \over
  2}(x\dot{y}-y\dot{x})+fy~. \lab{(26)} \ee
Note that $\hbar$ is introduced solely for dimensional reasons.
We thus see that the constraint condition at the classical level
introduced a new coordinate $y$, and the system equations are obtained: 
\be
m\ddot{x}+\ga \dot{x}=f ~,\ \  \
m\ddot{y}-\ga \dot{y}=0~. \lab{(27)}
\ee
The
$x$-system is an {\it open} system. In order to set up the canonical
formalism it is required to {\it close} the system; this is the r\^ole
of the $y$-system, which is the time-reversed copy of the
$x$-system. The $\{x-y\}$ system is thus a closed system.
We also remark that the exact expression for the imaginary part of the
action  reads~\cite{Srivastava:1995yf,DifettiBook}
\be \label{30c} {\rm Im}{\cal S}[x,y]=
\frac{1}{2\hbar}\int_{t_i}^{t_f}\int_{t_i}^{t_f}dt~ds
\,N(t-s)~y(t)~y(s) ~, \ee
where $N(t-s)$ denotes the quantum noise in the fluctuating random
force given by the Nyquist theorem~\cite{Srivastava:1995yf}. From Eq.~(\ref{30c}) we see  that nonzero $y$ yields an
``unlikely process'' in the classical limit ``$\hbar \rightarrow 0$'',
in view of the large imaginary part of the action. At quantum level,
instead, nonzero $y$ may allow quantum noise effects arising from the
imaginary part of the action~\cite{Srivastava:1995yf}.  We thus see that the second
sheet cannot be neglected: in the perturbative approach one may drop
higher order terms in the action functional expansion, since they
correspond to unlikely processes at the classical level. However,
these terms may be responsible for quantum noise corrections
and thus, in order to not preclude the quantization effects, one
should keep them.

Let us unveil now the  relation between the
two-sheeted space in the NCSG construction and the gauge structure of
the theory.
Consider the equation of the classical one-dimensional damped
harmonic oscillator
\be
m \ddot x + \gamma \dot x + k x  = 0~, \label{2.1a}
\ee
with time independent $m$, $\gamma$ and $k$, which is a simple
prototype of open systems. 
In the canonical formalism for open
systems, the doubling of the degrees of freedom is required in such a
way  as to complement the given open system with its
time-reversed image, playing the r\^ole of the ``bath", thus obtaining a globally closed system for which
the Lagrangian formalism is well defined. Thus we consider the oscillator in the {\it doubled} $y$ coordinate
\be
m \ddot y - \gamma \dot y + k y  = 0 ~.\label{2.1b}
\ee
The system of the oscillators Eq.~(\ref{2.1a}) and  Eq.~(\ref{2.1b})
is then a closed system described by the Lagrangian density
Eq.~(\ref{(26)}) where we put $f = k x$.  The canonically conjugate
momenta $p_{x}$ and $p_{y}$ can now be introduced as customary. 
Let us use the coordinates ${{x_1}(t)}$ and ${{x_2}(t)}$:
$x_{1}(t) = (x(t) + y(t))/\sqrt{2}$ and $x_{2}(t) =
(x(t) - y(t))/\sqrt{2}$.
The motion equations are 
\be
m \ddot x_1 + \gamma \dot x_2 + k x_1  = 0~, \label{2.16} \qquad
m \ddot x_2 + \gamma \dot x_1 + k x_2   = 0~,
\ee
and $\, p_{1} = m {\dot x}_{1} + (1/2) \ga {x_2}$ ; $p_{2} = - m
{\dot x}_{2} - (1/2) \ga {x_1} \,$.  
Following
Refs.~\cite{Tsue:1993nz,Blasone:1996yh,Celeghini:1992a,Celeghini:1993a}
we can now put $B \equiv \gamma \,{c/{e}}$, ~${\epsilon}_{ii} = 0$, ~${\epsilon}_{12}
  = - {\epsilon}_{21} = 1$ and introduce the vector potential as
\be A_i = {B\over 2} \epsilon_{ij} x_j ~~~~(i,j = 1,2)~.\label{2.21}
\ee
The Lagrangian can be written then
in the familiar form
\bea L &=& {1 \over 2m} (m{\dot x_1} + {e_1
  \over{c}} A_1)^2 - {1 \over 2m} (m{\dot x_2} + {e_2 \over{c}} A_2)^2 - {e^2\over 2mc^2}({A_1}^2 + {A_2}^2) -
e\Phi \label{2.24i}~,
\eea
which describes two
particles with opposite charges $e_1 = - e_2 = e$ in the (oscillator)
potential $\Phi \equiv (k/2e)({x_1}^2 - {x_2}^2) \equiv {\Phi}_1
- {\Phi}_2$ with $ {\Phi}_i \equiv (k/2/e){x_i}^{2}$ and in the
constant magnetic field $\boldvec{B}$ defined as $\boldvec{B}=
\boldvec{\nabla} \times \boldvec{A} = - B \boldvec{{\hat 3}}$.
Remarkably, we have the Lorentzian-like (pseudoeuclidean) metric in
Eq.~(\ref{2.24i}).   The ``minus" sign, implied by the doubling of the degrees of freedom, is crucial in our
derivation (and in the NCSG construction).

In conclusion, the doubled coordinate, e.g., $x_2$ acts as the gauge
field component $A_1$ to which the $x_1$ coordinate is coupled, and
{\sl vice versa}. The energy dissipated by one of the two systems is
gained by the other one and viceversa, in analogy to what happens in
standard electrodynamics as observed at the end of Section~I.  The
interpretation is recovered of the gauge field as the bath or
reservoir in which the system is
embedded~\cite{Celeghini:1992a,Celeghini:1993a}. The gauge structure
thus appears intrinsic to the doubling procedure.

Such a conclusion can be also reached in the case of a
fermion field. For brevity we do not report here the discussion for the fermion case, which can be found in \cite{Celeghini:1992a,Celeghini:1993a,PRD}. We only observe that, considering as an example 
the Lagrangian
of the massless free Dirac field $L=-\overline{\psi}\gamma^{\mu}\partial_{\mu}\psi$, 
the  field algebra is doubled by introducing
the fermion tilde-field $\tilde{\psi}(x)$ and 
the Lagrangian  is rewritten as
\be \hat L = L - {\tilde L} = - \overline{\psi}
\gamma^{\mu}\partial_{\mu}\psi + \overline{\tilde {\psi}}
\gamma^{\mu}\partial_{\mu} \tilde{\psi}~. \label{(7)} \ee
The tilde-system is a ``copy"
(with the same spectrum and couplings) of the $\psi$-system.
The Hamiltonian for the system is of the form ${\hat H} = H - {\tilde
  H}$. The key point is that 
the matrix
elements of the Lagrangian Eq.~(\ref{(7)}) in a conveniently introduced space of states ${\cal H}_{\theta} \equiv \{|0 (\theta) \rangle\}$, where $|0 (\theta) \rangle$ denotes the ground state (see \cite{Celeghini:1992a,Celeghini:1993a,DifettiBook,PRD} ), as well
as of a more general Lagrangian than the simple one presently
considered, are invariant under the simultaneous local gauge
transformations of $\psi$ and $\tilde \psi$. The label $\theta$ in ${\cal H}_{\theta}$ denotes the angle of a Bogoliubov transformation (see below).
The tilde term $\overline{\tilde {
    \psi}}\gamma^\mu\partial_{\mu}\tilde{\psi}$  transforms in
such a way to compensate the local gauge transformation of the
$\psi$ kinematic term, i.e. $\overline {\tilde \psi} (x) \gamma^{\mu}\partial_{\mu} \tilde \psi
(x) \to \overline {\tilde \psi} (x) \gamma^{\mu} \partial_{\mu} \tilde
\psi (x) + g \partial^{\mu}\alpha(x) \tilde{j}_{\mu}(x)$. 
This suggests  to introduce the  field $A_{\mu}^\prime$ by
\be gj^{\bar \mu} (x) A_{\bar \mu}^\prime(x)\cong \overline {\tilde \psi}
(x) \gamma^{\bar \mu}
\partial_{\bar \mu} \tilde \psi (x)~,~~~
\bar \mu=0,1,2,3~. \label{(16)} \ee
where the bar over ${\mu}$ means no summation
over repeated indices. The symbol $\cong$ denotes equality among matrix elements in ${\cal H}_{\theta}$. Thus we find that  $A_{\mu}^\prime$
transforms as $A_{\mu}^\prime(x) \to A_{\mu}^\prime(x) +
\partial_{\mu}\alpha(x)$, 
and one may 
identify, in ${\cal H}_{\theta}$, $A_{\mu}^\prime$ with the
conventional U(1) gauge vector field.

One can also show that the
variations of the gauge field tensor $F_{\mu \nu}^\prime$ have their
source in the current $\tilde j_{\mu}$, which suggests that the tilde
field plays the r\^ole of the ``bath" or ``reservoir". Such an interpretation in
terms of a reservoir, may thus be extended also to the gauge field
$A_{\mu}^\prime$, which indeed acts in a way to ``compensate" the
changes in the matter field configurations due to the local gauge
freedom.

Finally, it can be shown that in
the formalism of the algebra doubling a relevant r\^ole is played by
the noncommutative $q$-deformed Hopf algebra~\cite{Celeghini:1998a},
pointing to a deep physical meaning of the noncommutativity in this
construction. Indeed, the map ${\cal A} \, \rightarrow \,{\cal A}_1
\otimes {\cal A}_2$ is just the Hopf coproduct map
${\cal A} \, \rightarrow \, {\cal A} \otimes \mathds{1} + \mathds{1}
\otimes {\cal A} \equiv \, {\cal A}_1 \otimes {\cal A}_2$ which
duplicates the algebra.  The Bogoliubov transformation of ``angle"
$\theta$ relating the fields $\psi (\theta; x)$ and ${\tilde \psi}
(\theta; x)$ to $\psi (x)$ and ${\tilde \psi} (x)$, is known to be
obtained by convenient combinations of the {\it deformed} coproduct
operation of the form $\Delta a^{\dag}_q=a^{\dag}_q\otimes q^{1/2} +
q^{-1/2}\otimes a^{\dag}_q$, where $q \equiv q (\theta)$ is the
deformation parameters and $a^{\dag}_q$ are the creation operators in
the $q$-deformed Hopf algebra~\cite{Celeghini:1998a}. These deformed
coproduct maps are noncommutative and the deformation parameter is
related
to the condensate content of
$|0 (\theta) \rangle$.  It is interesting to observe
that the $q$-derivative is a finite difference derivative~\cite{Celeghini:1998a}, which has
to be compared with the fact that in the NCSG construction the
derivative in the discrete direction is a finite difference quotient.

A relevant point is that the deformation parameter {\it labels} the
$\theta$-representations $\{|0 (\theta) \rangle\}$ and, for $\theta
\neq \theta'$, $\{|0 (\theta) \rangle\}$ and $\{|0 (\theta')
\rangle\}$ are unitarily inequivalent representations of the canonical
(anti-)commutation rules. This is a characteristic feature of quantum
field theory~\cite{DifettiBook,Umezawa:1982nv}. Its physical meaning
is that an order parameter exists, which assumes different
$\theta$-dependent values in each of the representations. In other
words, the deformed Hopf algebra structure induces the {\it foliation}
of the whole Hilbert space into physically inequivalent subspaces.

\section{Dissipation and quantization}\label{4}

By discussing classical, deterministic models, 't~Hooft  has
conjectured that, provided some specific energy conditions are met and
some constraints are imposed, loss of information might lead to a
quantum evolution \cite{'tHooft:1999gk,erice,'tHooft:2006sy}. In agreement
with 't~Hooft's conjecture, on the basis of the discussion in the previous Sections and following
Refs.~\cite{Blasone:2000ew,Blasone:2002hq}, we propose~\cite{PRD} that the NCSG classical construction carries {\it implicit} in
its feature of the doubling of the algebra the seeds of quantization.

We consider the classical damped harmonic
$x$-oscillator described by Eq.~(\ref{2.1a}) and its time--reversed
image Eq.~(\ref{2.1b}). 
The Casimir operator {\cal C} and the (second) $SU(1,1)$ generator $J_2$ 
are~\cite{Blasone:1996yh}
\bea\lab{ca} {\cal C} = \frac{1}{4 \Om m}\lf[ \lf(p_1^2  - p_2^2\ri)+
m^2\Om^2 \lf(x_1^2 -  x_2^2\ri)\ri]~, \qquad
J_2 = \frac{m}{2}\lf[\lf( {\dot x}_1 x_2 - {\dot x}_2
x_1 \ri) - {\Ga} r^2 \ri]
~,
\eea
where ${\cal C}$ is taken to be positive and
$$\Ga = {\ga\over 2 m}~,~ \Om = \sqrt{\frac{1}{m}
(k-\frac{\ga^2}{4m})}~,~ \mbox{with}~~ k >\frac{\ga^2}{4m}~.$$
The system's Hamiltonian can be written as \cite{Blasone:2000ew,Blasone:2002hq}
\bea \lab{pqham}
H &=& \sum_{i=1}^2p_{i}\, f_{i}(q)\,,
\eea
with $p_1 = {\cal C}$, $p_2 = J_2$, $f_1(q)=2\Om$, $f_2(q)=-2\Ga$, 
$\{q_{i},p_i\} =1$ and
the other Poisson brackets are vanishing.
The Hamiltonian Eq.~(\ref{pqham}) belongs to the class of Hamiltonians
considered by 't~Hooft. There, the $f_{i}(q)$ are nonsingular
functions of the canonical coordinates $q_{i}$ and the equations for
the $q$'s, namely $\dot{q_{i}} = \{q_{i}, H\} = f_{i}(q)$), are
decoupled from the conjugate momenta $p_i$. A complete set of observables, called {\em beables}, then exists, which
Poisson commute at all times.  The meaning of this is that
the system admits a deterministic description even when expressed in
terms of operators acting on some functional space of states
$|\psi\ran$, such as the Hilbert space~\cite{erice}. Such a description in terms of operators and Hilbert space, does not
imply {\em per se} quantization of the system.  Quantization is achieved only as a consequence of dissipation. 
The Hamiltonian is written as $H = H_{\rm \I} - H_{\rm \II}$, with
\bea
&&H_{\rm \I} = \frac{1}{2 \Om {\cal C}} (2 \Om {\cal C}
- \Ga J_2)^2 ~~,~~
H_{\rm \II} = \frac{\Ga^2}{2 \Om {\cal C}} J_2^2\,   \label{split}
\eea
and we impose the constraint $J_2
|\psi\ran = 0$, 
which defines
physical states and guaranties that $H$ is bounded from below.
We can then write
\bea \lab{17}
H |\psi\ran= H_{\rm \I} |\psi\ran=  2\Om {\cal C}|\psi\ran
= \left( \frac{1}{2m}p_{r}^{2} + \frac{K}{2}r^{2}\right) |\psi \ran \, ,
\eea
with $K\equiv m \Om^2$. We thus realize that $H_{\rm \I}$ reduces to the
Hamiltonian for the two-dimensional ``isotropic'' (or ``radial'')
harmonic oscillator $\ddot{r} + \Om^2 r =0 $.

The physical states are invariant under time-reversal ($|\psi(t)\ran =
|\psi(-t)\ran$) and periodical with period $\tau = 2\pi/\Omega$. 
Note that $ H_{\rm \I} = 2 \Om{\cal C} $ has the spectrum ${\cal
  H}^n_{\rm \I}= \hbar \Om n$, $n = 0, \pm 1, \pm 2, ...$; since our
choice has been that ${\cal C}$ is positive, only positive values of
$n$ will be considered.

By exploiting the periodicity of the physical states
$|\psi\rangle$ and following Ref.~\cite{Berry}, one obtains
$$\frac{ \langle \psi(\tau)| H |\psi(\tau) \rangle }{\hbar} \tau -
\phi = 2\pi n~~,~~ n = 0, 1, 2, \ldots~.$$  Using $\tau = 2
\pi/\Om$ and $\phi = \alpha \pi$, where $\al$ is a real constant, leads to
\bea\lab{spectrum}
{\cal H}_{\rm \I,e\!f\!f}^n \equiv
\langle \psi_{n}(\tau)| H |\psi_{n}(\tau) \rangle= \hbar
\Om \lf( n + \frac{\alpha}{2} \ri) ~.
\eea
The index $n$ signals the $n$ dependence of the
state and the corresponding energy.  ${\cal H}_{\rm
  \I,e\!f\!f}^n$ gives the effective $n$th energy level of the
system, namely the energy given by ${\cal H}_{\rm \I}^n$
corrected by its interaction with the environment. We conclude that
the dissipation term $J_2$ of the Hamiltonian is responsible for the
zero point ($n = 0$) energy: $E_{0} =(\hbar/2) \Om \alpha$. 
In QM the zero point energy is formally
due to the nonzero commutator of the canonically conjugate $q$ and $p$
operators: the zero point energy {\it is} the ``signature" of
quantization. Our discussion thus shows that dissipation manifests
itself as ``quantization". In other words, the (zero point) ``quantum
contribution" $E_0$ to the spectrum of physical states signals the
underlying dissipative dynamics.

Consider the defining relation for temperature in thermodynamics  (with $k_{\rm B} = 1$): $\pa S/\pa U = 1/T$.
Using $S \equiv (2 J_{2}/\hbar)$ and $U \equiv 2 \Om {\cal C}$,
Eq.~(\ref{pqham}) gives $T = \hbar \Ga$.  Provided  $S$ is
identified with the entropy, $\hbar \Ga$ can be regarded as the
temperature.  Thus,  the ``full Hamiltonian'' Eq.~(\ref{pqham})
plays  the r{\^o}le of the free energy ${\cal F}$, and $2
\Ga J_{2}$ represents  the heat contribution in $H$ (or
$\cal F$).  The fact that $2 J_{2}/\hbar$ behaves as
the entropy is not surprising since it controls the dissipative (thus
irreversible loss of information) part of the dynamics.

The thermodynamical picture outlined above is
also consistent with the results on the canonical quantization of
open systems in quantum field theory~\cite{Celeghini:1992yv}.

\section{The dissipative interference phase}
\label{5}

Dissipation implies the appearance of a ``dissipative
interference phase''~\cite{Sivasubramanian:2003xy}. This provides a relation between dissipation and noncommutative
geometry in the plane of the doubled coordinates and thus with the NCSG construction.

In the $ (x_+,x_-)$  plane (cf.
Section \ref{classical}), the components  of forward and
backward in time velocity $ v_\pm =\dot{x}_\pm$ are given by
\be v_{\pm }={{\partial  H} \over
\partial p_{\pm }} =\pm \, \frac{1}{m}\lf( p_\pm \mp \frac{\ga}{2}\,
x_\mp \ri) ~, \label{(9)} \ee
and they do not commute
\be [v_+,v_-]=i\hbar \,{\ga\over m^2}~.  \label{(10)} \ee
It is thus impossible to fix these velocities $v_+$ and $v_-$ as being
identical~\cite{Sivasubramanian:2003xy}.
By putting $m v_\pm =\hbar K_\pm$,   
a canonical set of conjugate position coordinates
$ (\xi_+,\xi_-)$ may be defined by $\xi_\pm = \mp L^2K_\mp$ so that
\be
\left[\xi_+,\xi_-  \right] = iL^2. \label{DP2} \ee
Equation (\ref{DP2}) 
characterizes the noncommutative geometry in the plane $(x_+,x_-)$.
In full generality, one can show \cite{Sivasubramanian:2003xy} that an Aharonov--Bohm-type phase interference
can always be associated with the noncommutative $(X,Y)$ plane where
\begin{equation} \label{1noncomm}
[X,Y]=iL^2~;
\end{equation}
$L$ denotes the geometric length scale in the
plane. Consider a particle moving in the plane along two paths, $ {\cal P}_1
$ and $ {\cal P}_2 $, starting and finishing at the same point, in a
forward and in a backward direction, respectively. Let $ {\cal A} $
denote the resulting area enclosed by the paths. The
phase interference $\vartheta$ may be be shown~\cite{Sivasubramanian:2003xy} to be given by
\begin{equation}
\vartheta = \frac{\cal A}{L^2}~. \label{IntPhase}
\end{equation}
and Eq.~(\ref{1noncomm}) in the noncommutative plane can be written as
\begin{equation}
[X,P_X]=i\hbar \ \ \ {\rm where} \ \ \ P_X=\left(\frac{\hbar
Y}{L^2}\right)~. \label{phase4}
\end{equation}
The quantum phase
interference between two alternative paths in the plane is thus
determined by the noncommutative length scale
$ L  $ and the enclosed area
$ {\cal A} $.

Notice that the existence of a phase interference  is
connected to the zero point fluctuations in the
coordinates; indeed Eq.~(\ref{1noncomm}) implies a zero point
uncertainty relation
$ (\Delta X) (\Delta Y) \ge L^2/2~. $

In the  dissipative case,
$L^2=\hbar / \ga$,
we conclude that the quantum
dissipative phase interference $\vartheta = {\cal A}/L^2 =
{\cal A} \ga/\hbar$ is associated with the two paths ${\cal P}_1$
and ${\cal P}_2$ in the noncommutative plane, provided $x_+ \neq x_-$.

\section{Conclusions}
\label{concl}

We have shown that the central ingredient in the
NCSG, namely the doubling of the algebra ${\cal A}={\cal A}_1\otimes
{\cal A}_2$ acting on the space ${\cal H} = {\cal H}_1\otimes {\cal
  H}_2$ is
related to dissipation and to the gauge
field structure.  Thus the two-sheeted geometry must  be
considered to be 
the construction leading to gauge fields, required to explain
the Standard Model.  By exploiting 't Hooft's conjecture,
according which loss of information within the framework of completely
deterministic dynamics, might lead to a quantum evolution, we have
argued that dissipation, implied by the algebra doubling, may lead to
quantum features. We thus suggest that the NCSG classical
construction carries {\it implicit} in the doubling of the algebra the
seeds of quantization.

We have shown that in Alain Connes' two-sheeted construction, the
doubled degree of freedom is associated with ``unlikely processes'' in
the classical limit, and it may thus be dropped  in
higher order terms in the perturbative expansion. However, since these higher order
terms are the ones responsible for quantum
corrections, the second sheet cannot be neglected if one does not want to preclude quantization effects.  In other words, the second sheet cannot
be neglected once the universe entered the radiation dominated
era. However, at the Grand Unified Theories scale, when inflation took
place, the effect of gauge fields is fairly shielded.

\end{document}